
\input amstex
\documentstyle{amsppt}
\NoRunningHeads
\TagsOnRight
\magnification=1200
\document
$$\split\qquad\qquad\qquad\qquad\qquad\qquad\qquad\qquad\qquad\qquad\qquad
\qquad &\text{LMU-TPW 92-31}\\ &\ \text{January 1993}\endsplit$$
\bigskip\bigskip\bigskip
$$\gather\text{
{\bf ON THE GEOMETRICAL STRUCTURE OF}}\\\text{{\bf COVARIANT ANOMALIES IN
YANG-MILLS THEORY}}\endgather$$
$$\gather\text{Gerald Kelnhofer\footnotemark"*"}\\
\text{Sektion Physik der Ludwig-Maximilians-Universit\"at}\\
\text{Theresienstr. 37, 8000 M\"unchen 2}\\\text{Germany}\endgather$$
\bigskip\bigskip\bigskip\bigskip
\head Abstract\endhead
Covariant anomalies are studied in terms of the theory of secondary
characteristic classes of the universal bundle of Yang-Mills theory.
A new set of descent equations is derived which contains the covariant
current anomaly and the covariant Schwinger term. The
counterterms relating consistent and covariant anomalies are determined. A
geometrical realization of the BRS/anti-BRS algebra is presented
which is used to understand the relationship between covariant anomalies in
different approaches.
\footnotetext"*)"{Erwin Schr\"odinger fellow, supported by "Fonds zur
F\"orderung der wissenschaftlichen Forschung in \"Osterreich", project
number: J0701-PHY}%
\par\newpage
\head 1. Introduction\endhead
\bigskip
There has been much progress in investigating the phenomenon of anomalies in
quantum field theory during the last years. Anomalies manifest
themselves in
two different ways, known as consistent and covariant anomalies.
Cohomological methods have been proved to be useful for analyzing the
mathematical structure of the consistent anomaly. Its algebraic origin can
be traced back to the Weil algebra structure of the BRS algebra [DTV]. The
BRS algebra is based on the multiplet $(A,c,F_A,\delta ,d_P)$, where $A$
is a connection
with curvature $F_A$ in a principal $G$-bundle $P$ over $n$-dimensional
space-time $M$, $c$ is the (Faddeev-Popov)
ghost, $\delta$ is the BRS operator and $d_P$ is the exterior derivative on
$P$.
The consistent anomalies are related to the BRS cohomology $\delta$ modulo
$d_P$ and are generated by the consistent descent equations [S, Zu].\par
The universal bundle construction of Atiyah and Singer [AS]
provides a geometrical framework to realize the BRS multiplet and the
corresponding BRS transformations. Application of the theory of
secondary characteristic forms [C] finally leads to the consistent
descent equations.\par
There have been various attempts to understand covariant anomalies from a
cohomological viewpoint [Ts, Z, AAG, K1]. An algebraic method has been
suggested
in [Ts] and [Z] to derive covariant anomalies such as the covariant current
anomaly [BZ] and the covariant commutator
anomaly (Schwinger term) [NT, K2] by transgression from a
Chern-Simons form.
The basic idea is to enlarge the usual BRS algebra by introducing a new
field $\bar c$, the anti-ghost, together with a nilpotent operator
$\bar\delta$, which is called anti-BRS operator. This BRS/anti-BRS algebra
provides an algebraic framework to formulate an integrability condition
for covariant anomalous terms of arbitrary ghost degree.
In [Ts, Z], a solution of this covariance condition was
obtained by introducing certain homotopy operators.
A geometrical description of the BRS/anti-BRS algebra has been
proposed in [Z]. But this attempt is not satisfactory since the global
meaning of the anti-ghost, being defined as part of a connection in the
universal bundle, and of the anti-BRS operator is not very obvious. \par
A different characterization of the covariant anomaly has been already
suggested by Bonora and Cotta-Ramusino [BC1]. Their considerations
are
based on a reformulation of the anomaly problem in the total space of the
given principal fibre bundle $P$. They found that the non-integrated
covariant current
anomaly belongs to a certain class in the $n$th local de-Rham cohomology
group of ghost degree one built with cocycles of the total space $P$ linear
in the ghost. One aim of this paper is to understand the relationship
between these different approaches. \par
We present a geometrical approach to the calculation of covariant
anomalies in terms of secondary characteristic classes of the universal
bundle. Thereby
a new set of descent equations is derived which contains the covariant
current anomaly and the covariant Schwinger term. The result of [BC1]
follows directly from our analysis.\par
In order to relate the different approaches we shall construct a geometrical
realization of the BRS/anti-BRS algebra on an appropriate $G$ bundle.
Ghost and anti-ghost turn out to be components of a canonical connection in
this bundle. Our formalism naturally extends to this generalized bundle.
Finally
we obtain a system of descent equations whose solutions satisfy the
covariance condition. The possibility to characterize covariant anomalies
in terms of local cohomology turns out to be a consequence of the proposed
formalism.
\par
This paper is organized as follows: In Sec. 2, we fix some notations for
the geometry of Yang-Mills fields. Then we review the universal bundle
construction of Atiyah and Singer and derive the consistent descent
equations. In Sec. 3 we present our geometrical approach and derive the
covariant descent equations. The relation between consistent and
covariant anomalies is investigated. Then we compare our formalism with
the results of [BC1]. An appropriate geometrical realization of the
BRS/anti-BRS multiplet is constructed in Sec. 4. Then
we discuss the relationship between the covariant anomaly in the
two different approaches.
\bigskip\bigskip
\head 2. The consistent descent equations\endhead
\bigskip
In this section we shall review the geometrical structure of consistent
anomalies based on the universal bundle construction [AS].
In order to fix notation we shall begin with a short review of the
geometry of Yang-Mills fields. \par
Let $(M,g)$ be a $n$-dimensional, compact, oriented
Riemannian manifold, $\pi _P\colon P\rightarrow M$ a principal $G$-bundle
over $M$, where the structure group $G$ is a compact Lie group and whose
Lie algebra will be denoted by $\frak g$. The principal right action is
$(p,g)\mapsto pg$. The space of (irreducible)
connections on $P$ will be denoted by $\Cal A$ and the gauge group $\Cal G$
is identified with the group of based bundle automorphisms (modulo the centre
of $G$) of $P$. So $\pi _{\Cal A}\colon\Cal A\rightarrow
\Cal M$ becomes a principal-$\Cal G$ bundle over the gauge orbit space
$\Cal M$ with principal action $\Cal R^{\Cal A}(A,u):=u^{\ast}A$ [MV].
Every $u\in\Cal G$ defines a section $\hat u$ of the
associated bundle $P\times _GG$, given by $u(p)=p\hat u(p)$. Associated to
$P$ by the adjoint action of $G$ in $\frak g$ is the
bundle $adP:=P\times _G\frak g$. The bundles $\wedge ^q
T^{\ast}M\otimes adP$ have canonical metrics $(,)$ induced by the metric $g$
and the Killing form on $\frak g$ and thus the space of $adP$-valued
$q$-forms
$$\Omega ^q(M,adP):=\Gamma (\wedge ^q T^{\ast}M\otimes adP)\tag2.1$$
has an inner product, given by
$$<\phi _1,\phi _2>:=\int_M(\phi _1,\phi _2).\tag2.2$$
Every $A\in\Cal A$ defines a covariant exterior derivative $d_A
\colon\Omega ^q(M,adP)\rightarrow\Omega ^{q+1}(M,adP)$. Its adjoint will be
denoted by $d_A^{\ast}$. The Lie algebra $Lie
\Cal G$ of $\Cal G$ can be identified with $\Omega ^0(M,adP)$ and the
fundamental vector field $Z^{\Cal A}$ with respect to $\Cal R^{
\Cal A}$ is given by $Z_{\xi}^{\Cal A}(A)=d_A\xi$, where
$\xi\in\Omega^0(M,adP)$.
Since $\Cal A$ is an affine space the tangent bundle is $T\Cal A=\Cal A\times
\Omega ^1(M,adP)$. Note that
$\Omega ^q(M,adP)$ is isomorphic with the space
of $ad$-equivariant, horizontal $\frak g$-valued $q$-forms on $P$, denoted
by $\Omega _{eq,h}^q(P,\frak g)$.\par
There exists a canonical connection $\alpha$ on the bundle $\Cal A\rightarrow
\Cal M$, given by
$$\alpha _A(\tau _A)=(d_A^{\ast}d_A)^{-1}d_A^{\ast}(\tau _A).\tag2.3$$
Thus every tangent vector $\tau _A\in T_A\Cal A$ can be split into a
horizontal and vertical part, written as
$$\tau _A=\tau _A^h+\tau _A^{\text{ver}}:=(1-d_A\alpha _A)\tau _A+d_A
\alpha _A\tau _A.\tag2.4$$
Each $\tau\in\Omega ^1(M,adP)$ defines the operator
$B_{\tau}\phi :=[\tau ,\phi ]$ with $\phi\in\Omega ^q(M,adP)$.
The curvature $\Cal F$ of $\alpha$ reads
$$\Cal F_A(\tau _1,\tau _2)=(d_A^{\ast}d_A)^{-1}(B_{\tau _1^h}^{\ast}
\tau _2^h-B_{\tau _2^h}^{\ast}\tau _1^h),\tag2.5$$
where $B_{\tau}^{\ast}$ denotes the adjoint of $B_{\tau}$. \par
Atiyah and Singer introduced the universal bundle with a
canonical connection in order to determine the characteristic
classes of the index bundle of the family of Dirac operators coupled to
gauge fields. The first Chern class turned out to
represent an obstruction to the existence of a non vanishing gauge invariant
determinant for fermions in the background of external Yang-Mills fields.
A detailed review of the relation between index theory and non-abelian
anomalies can be found in [Tr].
Now we shall
briefly recall the construction of the universal bundle.
\par
Consider the $G$-bundle $\Cal A\times P@>\pi >>\Cal A\times M$, denoted by
$\Cal Q$, with principal action $R_g(A,p):=(A,pg)$. There exists a free
right $\Cal G$-action on $\Cal Q$, given by $\Cal R_u(A,p):=(u^{\ast}A,
u^{-1}(p))$ so that the quotient $(\Cal A\times P)/\Cal G$ becomes a
principal $G$ bundle $\hat\Cal Q$ with base space $\Cal M\times M$. The
bundle $\hat\Cal Q$ is called the universal bundle of Yang-Mills theory.
We can summarize this construction by the commutative diagram
$$\CD\Cal A\times P @>q>>(\Cal A\times P)/\Cal G\\
@V\pi VV @V\hat\pi VV\\
\Cal A\times M @>\hat q>>\Cal M\times M\endCD\tag2.6$$
Obviously, one has $\hat q^{\ast}\hat\Cal Q\cong\Cal Q$.
The algebra $\Omega (\Cal A\times P)$ of differential forms on $\Cal A
\times P$ admits a bigrading $\Omega (\Cal A\times P)=\oplus _{i,j\geq 0}
\Omega ^{(i,j)}(\Cal A\times P)$. Thus a form $\phi\in\Omega ^{(i,j)}(\Cal A
\times P)$ will be denoted by $\phi ^{(i,j)}$.\par
There exists a canonical connection on $\Cal Q$, defined by
$$\gamma _{(A,p)}(\tau _A,X_p)=\gamma _{(A,p)}^{(0,1)}(X_p)+\gamma _{(A,p)}
^{(1,0)}(\tau _A):=A_p(X_p)+(\alpha _A(\tau _A))(p),\tag2.7$$
where $X_p\in T_pP,\ \tau _A\in T_A\Cal A$. It can be shown that $\gamma$ is
$\Cal G$-invariant. The vertical bundle $V^q(\Cal A
\times P)$ of the principal $\Cal G$-bundle $\Cal A\times P@>q>>(\Cal A
\times P)/\Cal G$ is given by
$$V^q(\Cal A\times P)=\bigcup _{(A,p)\in\Cal A\times P}\lbrace
(d_A\xi ,-Z_{\xi}(p))\in T_{(A,p)}(\Cal A\times P)\vert\ \xi\in Lie\Cal G
\rbrace ,\tag2.8$$
where $Z_{\xi}$ denotes the fundamental vector field on $P$ generated by
$\xi$. Since $\gamma\vert _{V^q(\Cal A\times P)}=0$, the connection $\gamma$
descends to a connection $\hat\gamma$ on the universal bundle, so that
$\gamma =q^{\ast}\hat\gamma$. It has been shown in [AS] that this connection
is universal for any family of connections on $P$ which is parametrized by
any compact space.\par
The curvature $\Omega$ of $\gamma$ can be calculated by the formula
$\Omega =d_{\Cal A\times P}\gamma +\frac{1}{2}[\gamma ,\gamma ]$. Here,
$d_{\Cal A\times P}=d_{\Cal A}+\hat d_P$, where
$\hat d_P:=(-1)^id_P\colon\Omega ^{(i,j)}(\Cal A\times P)\rightarrow
\Omega ^{(i,j+1)}(\Cal A\times P)$. Being a
two form on $\Cal A\times P$, $\Omega$ decomposes into three components
which are given by
$$\aligned \Omega _{(A,p)}^{(0,2)}(X_p^1,X_p^2) &=(F_A)_p(X_p^1,X_p^2)\\
\Omega _{(A,p)}^{(2,0)}(\tau _A^1,\tau _A^2) &=(d_{\Cal A}\alpha +
\frac{1}{2}[\alpha ,\alpha ])_{(A,p)}(\tau _A^1,\tau _A^2)=(\Cal F_A(
\tau _A^1,\tau _A^2))(p)\\
\Omega _{(A,p)}^{(1,1)}(\tau _A,X_p) &=(d_{\Cal A}\gamma ^{(0,1)}-d_P
\gamma ^{(1,0)}-[\gamma ^{(0,1)},\gamma ^{(1,0)}])_{(A,p)}(\tau _A,X_p)
\\ &=(\tau _A)_p-
(d_A\alpha _A(\tau _A))_p(X_p)=(\tau _A^h)_p(X_p),\endaligned\tag2.9$$
where we used that
$$(d_{\Cal A}\gamma ^{(0,1)})_{(A,p)}(\tau _A,X_p)=\frac{d}{dt}\vert _{t=0}
\gamma _{(A+t\tau _A,p)}^{(0,1)}(X_p)=(\tau _A)_p(X_p).\tag2.10$$
By construction, $\Omega$ descends to the curvature $\hat\Omega$ of $\hat
\gamma$.\par
The BRS relations can now be obtained in the following way: Let
$$\bar i_A\colon\Cal G\times P\hookrightarrow\Cal A\times P,\qquad\bar i_A
(u,p)=(i_A\times id_P)(u,p)=(u^{\ast}A,p)\tag2.11$$
be the embedding of a gauge orbit through $A\in\Cal A$. We denote the Maurer
Cartan form on $\Cal G$ by $\Theta\in\Omega ^1(\Cal G,Lie\Cal G)$. Any
tangent
vector $\Cal Y_u\in T_u\Cal G$ can be written in the form $\Cal Y_u=
\Cal Y_{\Theta _u(\Cal Y_u)}^{\text{left}}(u)$, where $\Cal Y_{\Theta _u(
\Cal Y_u)}^{\text{left}}$ denotes the left invariant vector field on $\Cal G$
induced by $\Theta _u(\Cal Y_u)\in Lie\Cal G$. Then the restriction of
$\gamma$ to the gauge orbit through $A\in\Cal A$ reads
$$(\bar i_A^{\ast}\gamma )_{(u,p)}(\Cal Y_u,X_p)=(\bar i_A^{\ast}\gamma )
_{(u,p)}^{(0,1)}(X_p)+(\bar i_A^{\ast}\gamma )_{(u,p)}^{(1,0)}(\Cal Y_u)=
(u^{\ast}A)_p(X_p)+(\Theta _u(\Cal Y_u))(p).\tag2.12$$
Furthermore the restriction of the curvature gives the result
$$(\bar i_A^{\ast}\Omega )_{(u,p)}((\Cal Y_u^1,X_p^1),(\Cal Y_u^2,X_p^2))=
(u^{\ast}F_A)_p(X_p^1,X_p^2),\tag2.13$$
which is the so-called "Russian formula" [S]. There is a one to
one
correspondence between this formula and the BRS relations.
In fact,
setting $\bar A:=(\bar i_A^{\ast}\gamma )^{(0,1)}$
and $\bar F_A=\bar i_A^{\ast}\Omega$ we obtain from (2.9) and (2.13)
$$\aligned \bar F_A &=F_{\bar A}=d_P\bar A+\frac{1}{2}[\bar A,\bar A]\\ 0 &=
d_{\Cal G}
\Theta +\frac{1}{2}[\Theta ,\Theta ]\\ 0 &=d_{\Cal G}\bar A-d_{\bar A}
\Theta .
\endaligned\tag2.14$$
On the other hand, the Bianchi identity for $\bar i_A^{\ast}\Omega$ yields
$$d_{\Cal G}(\bar F_A)+[\Theta ,\bar F_A]=0.\tag2.15$$
These are the BRS relations [S], where $d_{\Cal G}$
and $\Theta$ are identified with the BRS operator and the ghost respectively.
(The minus in the last equation of (2.14) occurs because in our
convention $d_{\Cal G}$ and $d_P$ commute with each other.) Thus we have
explicitly verified that the bundle $\Cal Q$ provides an appropriate
geometrical framework to realize the BRS multiplet.\par
Let $I^m(G)$ denote the space of ad-invariant, symmetric multilinear real
valued functions on $\frak g$ of degree $m$. For a given $Q\in I^m(G)$ with
$m>\frac{n}{2}$, the consistent descent
equations can be derived from the transgression formula [C]
$$Q(\Omega )=d_{\Cal A\times P}TQ(\gamma ),\tag2.16$$
where $TQ(\gamma )=m\int _0^1dt\ Q(\gamma ,\Omega _t,\ldots ,\Omega _t)$
with $\Omega _t=t\Omega +\frac{t^2-t}{2}[\gamma ,\gamma ]$. Explicitly, the
components of $\Omega _t$ are given by
$$\aligned &\Omega _{t\ (A,p)}^{\ \ (0,2)}(X_p^1,X_p^2)=((F_A)_t)_p
(X_p^1,X_p^2)=(tF_A+\frac{t^2-t}{2}[A,A])_p(X_p^1,X_p^2)\\
& \Omega _{t\ (A,p)}^{\ \ (1,1)}(\tau _A,X_p)=(t\Omega _{(A,p)}^{(1,1)}+(t^2-
t)[\alpha _A,A])(\tau _A,X_p)\\ & \Omega _{t\ (A,p)}^{\ \ (2,0)}(\tau _A^1,
\tau _A^2)=(\Cal F_t)_A(\tau _A^1,\tau _A^2)(p)=(t\Cal F+
\frac{t^2-t}{2}[\alpha ,\alpha ])_A(\tau _A^1,\tau _A^2)(p)\endaligned
\tag2.17$$
With respect to the product structure of $\Omega (\Cal A\times P)$, (2.16)
gives
$$Q(\Omega )^{(k,2m-k)}=d_{\Cal A}TQ(\gamma )^{(k-1,2m-k)}+\hat d_PTQ(
\gamma )^{(k,2m-k-1)}\qquad 0\leqq k\leqq 2m,\tag2.18$$
The restriction of (2.18) to a gauge orbit through $A\in\Cal A$ finally
leads to the well known consistent descent equations [S, Zu]
$$\aligned Q(u^{\ast}F_A) &=d_P
TQ(u^{\ast}A)^{(0,2m-1)}=d_PTQ(\bar i_A^{\ast}\gamma )^{(0,2m-1)}\\ 0 &=
d_{\Cal G}TQ(\bar i_A^{\ast}\gamma )^{(k-1,2m-k)}+\hat d_PTQ(\bar i_A^{
\ast}\gamma )^{(k,2m-k-1)}\qquad 1\leqq k\leqq 2m.\endaligned\tag2.19$$
The relation for $k=2$ is the Wess-Zumino consistency condition [WZ].
Explicitly,
$$\multline TQ(\bar i_A^{\ast}\gamma )^{(1,2m-2)}\vert _{u=id_{\Cal G}}(\xi )
\\=
m\int _0^1dt\lbrace Q(\xi ,(F_A)_t,\ldots ,(F_A)_t)+(m-1)\frac{t^2-t}{2}Q(A,
[A,\xi ],(F_A)_t,\ldots ,(F_A)_t)\rbrace ,\endmultline\tag2.20$$
gives the non-integrated consistent current
anomaly [ZWZ] and $TQ(\bar i_A^{\ast}
\gamma )^{(2,2m-3)}\vert _{u=id_{\Cal G}}$ may be identified with the
non-integrated consistent Schwinger term [F, M]. \par
For $n=2m-2$, one finds that $Q(\Omega )^{(0,n+2)}=0$ for
dimensional reasons and therefore $TQ(\bar i_A^{\ast}\gamma )^{(0,n+1)}$ is
$d_P$ closed on $\Cal G\times P$, inducing the de Rham cohomology class
$$[TQ(\bar i_A^{\ast}\gamma )^{(0,n+1)}]\in H_{d_P}^{(0,n+1)}(\Cal G\times P,
\Bbb R).\tag2.21$$
This result has been firstly obtained in [BC1].\par
So far we have expressed the consistent anomalous terms by differential
forms on the space $\Cal G\times P$. It is often convenient to express them
in terms of forms on the base $\Cal G\times M$. So we choose a fixed
connection $a\in\Cal A$ and extend it to a connection
$$\gamma _{(A,p)}^a(\tau _A,X_p):=a_p(X_p)\tag2.22$$
in the $G$-bundle $\Cal A\times P\rightarrow\Cal A\times M$. Its
curvature $\Omega ^a$ is given
by
$\Omega ^a=F_a$ and therefore $Q(\Omega ^a)=Q(F_a)=0$ because of dimensional
reasons. Following [MSZ], we replace (2.16) by
$$Q(\Omega )=Q(\Omega )-Q(\Omega ^a)=d_{\Cal A\times P}TQ(\gamma ,\gamma ^a),
\tag2.23$$
with $TQ(\gamma ,\gamma ^a)=m\int _0^1dtQ(\gamma -\gamma ^a,\Omega _t^{
\prime},\ldots ,\Omega _t^{\prime})$ where $\Omega _t^{\prime}$ is the
curvature of the connection $\gamma _t=t\gamma +(1-t)
\gamma ^a$. The transgression form $TQ(\gamma ,\gamma ^a)$ is basic and
thus projects to a $(2m-1)$-form $\overline{TQ(\gamma ,\gamma ^a)}$ on
$\Cal A\times M$.\par
Applying the pullback $\bar i_A^{\ast}$ to (2.23), one is left with
$$\aligned Q(u^{\ast}F_A) &=d_PTQ(\bar i_A^{\ast}\gamma ,\bar i_A^{\ast}
\gamma ^a)^{(0,2m-1)}\\ 0 &=
d_{\Cal G}TQ(\bar i_A^{\ast}\gamma ,\bar i_A^{\ast}\gamma ^a)
^{(k-1,2m-k)}+
\hat d_PTQ(\bar i_A^{\ast}\gamma ,\bar i_A^{\ast}\gamma ^a)^{(k,2m-k-1)}
\qquad 1\leq k\leq 2m.\endaligned\tag2.24$$
Since the following identity holds
$$TQ(\gamma ,\gamma ^a)=TQ(\gamma )-TQ(\gamma ^a)+d_{\Cal A\times P}S_Q(
\gamma ,\gamma ^a),\tag2.25$$
where
$$S_Q(\gamma ,\gamma ^a)=m(m-1)\int _0^1dt_1\int _0^1dt_2\ t_2Q(\gamma -
\gamma ^a,\gamma _{t_1},(\Omega )_{t_2}^{(t_1)},\ldots ,(\Omega )_{t_2}
^{(t_1)}),\tag2.26$$
with $\gamma _{t_1}=t_1\gamma +(1-t_1)\gamma ^a$ and $(\Omega )_{t_2}
^{(t_1)}=t_2d_{\Cal A\times P}\gamma _{t_1}+\frac{t_2^2}{2}[\gamma _{t_1},
\gamma _{t_1}]$, we can explicitly verify that $TQ(\bar i_A^{\ast}
\gamma ,
\bar i_A^{\ast}\gamma ^a)^{(k,2m-k-1)}$ and $TQ(\bar i_A^{\ast}\gamma )^{(
k,2m-k-1)}$ represent the same cohomology class with respect to
the cohomology $d_{\Cal G}$ modulo $d_P$, i.e. they are cohomologous with
respect to the BRS-cohomology.\par
Let $N$ be a $(2m-k-1)$-dimensional, submanifold of $M$ without
boundary, then
the integrated consistent anomaly $\frak A^{(k)}$ of ghost degree $k$ is
identified with
$$\split\frak A^{(k)}(A,\xi _1,\ldots ,\xi _k) &:=
\int _N\overline{TQ(\bar i_A^{\ast}\gamma ,\bar i_A^{\ast}\gamma ^a)}
^{(k,2m-k-1)}\vert _{u=id_{\Cal G}}(\xi _1,\ldots ,\xi _k)\\ &=(i_{Z_{\xi _1}
^{\Cal A}}\cdots i_{Z_{\xi _k}^{\Cal A}}\int _N\overline{TQ(\gamma ,\gamma
^a)}^{(k,2m-k-1)})(A).\endsplit\tag2.27$$
It defines, for variable $A\in\Cal A$, a class of degree $k$ in the Lie
algebra cohomology of $Lie\Cal G$ with values in $C^{\infty}(\Cal A)$ [BC2].
\par
\bigskip\bigskip
\head 3. The covariant descent equations\endhead
\bigskip
In this section we want to show that the bundle $\Cal Q$ also provides an
appropriate framework to formulate descent equations for covariant anomalies.
\par
We define a connection on $\Cal Q$ by
$$(\gamma _0)_{(A,p)}(\tau _A,X_p):=A_p(X_p).\tag3.1$$
It is easily shown that this connection is $\Cal G$-invariant. However, it
does not induce a connection on $\hat\Cal Q$. The curvature
$\Omega _0$ of $\gamma _0$ decomposes into the three components
$$\aligned &\Omega _{0}^{\ \ (2,0)}=0\\ &
\Omega _{0\ (A,p)}^{\ \ (1,1)}(\tau _A,X_p)=(d_{\Cal A}A)_p(\tau _A,X_p)
=(\tau _A)_p(X_p)\\ &
\Omega _{0\ (A,p)}^{\ \ (0,2)}(X_p^1,X_p^2)=(F_A)_p(X_p^1,X_p^2).\endaligned
\tag3.2$$
Now we want to derive descent equations for the connection $\gamma _0$. Let
$Q\in I^m(G)$, then the equations
$d_{\Cal A\times P}Q(\Omega _0)=0$ and $Q(\Omega _0)=d_{\Cal A\times P}TQ(
\gamma _0)$ imply
$$\aligned 0 &=d_{\Cal A}Q(\Omega _0)^{(k-1,2m-k+1)}+\hat d_PQ(\Omega _0)
^{(k,2m-k)}
\qquad 0\leqq k\leqq m\\Q(\Omega _0)^{(k,2m-k)} &=d_{\Cal A}TQ(\gamma _0)^{
(k-1,2m-k)}+\hat d_PTQ(\gamma _0)^{(k,2m-k-1)}\qquad 0\leqq k\leqq m.
\endaligned\tag3.3$$
The components of $(
\Omega _0)_t:=t\Omega _0+\frac{t^2-t}{2}[\gamma _0,\gamma _0]$ are
$$\aligned &
(\Omega _0)_{t}^{\ \ (2,0)}=0,\\ & (\Omega _0)_{t\ (A,p)}^{\ \ (1,1)}
(\tau _A,X_p)=td_{\Cal A}A (\tau _A,X_p)=t(\tau _A)_p(X_p)\\
& (\Omega _0)_{t\ (A,p)}^{\ \ (0,2)}(X_p^1,X_p^2)=((F_A)_t)_p(X_p^1,X_p^2).
\endaligned\tag3.4$$
Thus it is evident from (3.2) and (3.4) that $Q(\Omega _0)^{(k,2m-k)}=0$
for $k\geq m+1$ and $TQ(\gamma _0)^{(k,2m-k-1)}=0$ for $k\geq m$.
However, we
should mention that there is no analog of the Russian formula for
$\Omega _0$. In fact, the restriction of $\Omega _0$ to a gauge orbit yields
$$\aligned & (\bar i_A^{\ast}\Omega _0^{(2,0)})=0\\
& (\bar i_A^{\ast}
\Omega _0^{(1,1)})_{(u,p)}(\Cal Y_u,X_p) =(d_{u^{\ast}A}\Theta _u(
\Cal Y_u))_p(X_p)\\ & (\bar i_A^{\ast}\Omega _0^{(0,2)})_{(u,p)}
(X_p^1,X_p^2)=(u^{\ast}F_A)_p(X_p^1,X_p^2),\endaligned\tag3.5$$
where the BRS relations (2.14) have been used.\par
If we define the secondary characteristic form $TQ(\gamma ,\gamma _0)\in
\Omega ^{2m-1}(\Cal A\times P)$ by
$$Q(\Omega )-Q(\Omega _0)=d_{\Cal A\times P}TQ(\gamma ,\gamma _0),\tag3.6$$
the corresponding set of descent equations reads
$$Q(\Omega )^{(k,2m-k)}-Q(\Omega _0)^{(k,2m-k)}=d_{\Cal A}TQ(\gamma ,
\gamma _0)^{(k-1,2m-k)}+\hat d_PTQ(\gamma ,\gamma _0)^{(k,2m-k-1)}.\tag3.7$$
The curvature $\Omega _t^{\prime}$ of the interpolating connection
$t\gamma +(1-t)\gamma _0$ has the components
$$\aligned &\Omega _{t\ (A,p)}^{\prime\ (2,0)}(\tau _A^1,\tau _A^2) =(
\Cal F_t)_A(\tau _A^1,\tau _A^2)(p)\\
& \Omega _{t\ (A,p)}^{\prime\ (1,1)}(\tau _A,X_p)=(d_{\Cal A}A-td_A
\alpha _A)(\tau _A,X_p)=(\tau _A-td_A\alpha _A(\tau _A))_p(X_p)\\
&\Omega _{t\ (A,p)}^{\prime\ (0,2)}(X_p^1,X_p^2) =(F_A)_p(X_p^1,X_p^2).
\endaligned\tag3.8$$
Finally, the restriction to a gauge orbit in $\Cal A$ leads to the
generalized descent equations
$$-Q(\bar i_A^{\ast}\Omega _0)^{(k,2m-k)}=d_{\Cal G}TQ(\bar i_A^{\ast}
\gamma ,\bar i_A^{\ast}\gamma _0)^{(k-1,2m-k)}+\hat d_PTQ(\bar i_A^{\ast}
\gamma ,\bar i_A^{\ast}\gamma _0)^{(k,2m-k-1)},\tag3.9$$
where $1\leq k\leq 2m$. In order to prove the relevance of our construction
for covariant anomalies we calculate, using (3.8)
$$TQ(\bar i_A^{\ast}\gamma ,\bar i_A^{\ast}\gamma _0)^{(1,2m-2)}\vert _{u=id_
{\Cal G}}=mQ(\Theta ,F_A,\ldots ,F_A),\tag3.10$$
which is the non-integrated covariant anomaly [BZ]. For $k=2$,
we obtain
$$TQ(\bar i_A^{\ast}\gamma ,\bar i_A^{\ast}\gamma _0)^{(2,2m-3)}\vert _{u=
id_{\Cal G}}=\frac{m(m-1)}{2}Q(\Theta ,d_A\Theta ,F_A,\ldots ,F_A).\tag3.11$$
This result agrees with previous calculations of the non-integrated
covariant Schwinger term [Ts, AAG].
Thus we have derived a new set of descent equations which contains the
covariant current anomaly and the covariant Schwinger term.
The differential forms
$$Q(\Omega _0)_{(A,p)}^{(k,2m-k)}=\binom mk Q(\undersetbrace
\text{$k$ times}\to{d_{\Cal A}A,\ldots ,d_{\Cal A}A},
\undersetbrace\text{$m-k$ times}\to{F_A,\ldots ,F_A})(p)
\tag3.12$$
vanish for $k\leq m-1$ if and only if $A$ is a flat connection. In general,
the forms
$TQ(\bar i_A^{\ast}\gamma ,\bar i_A^{\ast}\gamma _0)
^{(k-1,2m-k)}$ do not define cohomology classes $d_{\Cal G}$ modulo $d_P$.
Although this fact is well known for the cases $k=1,2$, we have explicitly
identified its geometrical origin.\par
Now we want to study the properties of $TQ(\gamma ,\gamma _0)$ under gauge
transformations. Therefore let $\overline{TQ(\gamma ,\gamma _0)}$ denote
the projection of
the transgression form $TQ(\gamma ,\gamma _0)$ onto $\Cal A\times M$. Since
$TQ(\gamma ,\gamma _0)$ is $\Cal G$ invariant with respect to $\Cal R$
, $\pi\circ\Cal R_u=(\Cal R_u^{\Cal A}\times id_M)\circ\pi$ implies
$$(\Cal R_u^{\Cal A}\times id_M)^{\ast}\overline{TQ(\gamma ,\gamma _0)}=
\overline{TQ(\gamma ,\gamma _0)}.\tag3.13$$
Since $\pi\circ (\Cal R_u^{\Cal A}\times id_P)=(\Cal R_u^{\Cal A}
\times id_M)\circ\pi$, one finally obtains
$$(\Cal R_u^{\Cal A}\times id_P)^{\ast}TQ(\gamma ,\gamma _0)=TQ(\gamma ,
\gamma _0).\tag3.14$$
Let $\wedge Lie\Cal G^{\ast}$ denote the exterior algebra of the dual of
the gauge algebra and let $ad^{\ast}$ denote the coadjoint action of
$\Cal G$ in $Lie\Cal G^{\ast}$. We define the map
$$\aligned & h^{(k)}\colon\Omega ^{(k,2m-k-1)}(\Cal A\times P)\rightarrow
\Omega ^{(0,2m-k-1)}(\Cal A\times P,\wedge ^{k}Lie\Cal G^{\ast})\\
& (h^{(k)}\phi ^{(k,2m-k-1)})_{(A,p)}(X_1,\ldots ,X_{2m-k-1})(\xi _1,
\ldots ,\xi _{k}):=\\ & \phi _{(A,
p)}^{(k,2m-k-1)}(Z_{\xi _1}^{\Cal A}(A),\ldots ,
Z_{\xi _
{k}}^{\Cal A}(A),X_1,\ldots ,X_{2m-k-1}),\endaligned\tag3.15$$
where $X_i\in T_pP$ and $\xi _i\in Lie\Cal G$.
The form $\chi ^{(k,2m-k-1)}:=h^{(k)}TQ(\gamma ,\gamma _0)^{(k,2m-k-1)}$ is
basic with projection $\bar \chi ^{(k,2m-k-1)}$. We have now the following
\proclaim{3.16 Proposition} $(\Cal R_u^{\Cal A}\times id_P)^{\ast}\chi
^{(k,2m-k-1)}=ad^{\ast}(\hat u^{-1})\chi ^{(k,2m-k-1)}$.\endproclaim
\demo{Proof}
Using (3.14), one finds
$$\split (( &\Cal R_u^{\Cal A}\times id_P)^{\ast}\chi ^{(k,2m-k-1)})_{(A,p)}
(X_1,\ldots ,X_{2m-k-1})(\xi _1,\ldots ,
\xi _{k})\\ &=(TQ(\gamma ,\gamma _0)^{(k,2m-k-1)})_{(u^{\ast}A,p)}(T_A
\Cal R_u^{\Cal A}Z_{ad(u)\xi _1}^{\Cal A},\ldots ,T_A
\Cal R_u^{\Cal A}Z_{ad(u)\xi _{k}}^{\Cal A},X_1,\ldots ,
X_{2m-k-1})\\ &=(\chi ^{(k,2m-k-1)})_{(u^{\ast}A,
p)}(X_1,\ldots ,X_{2m-k-1})(ad(u)\xi _1,\ldots ,ad(u)\xi _{k}).\endsplit$$
\enddemo
Let $N$ be a $(2m-k-1)$-dimensional, submanifold of $M$ without
boundary then we define the covariant anomaly $\tilde\frak A^{(k)}$ of
ghost degree $k$ by
$$\split\tilde\frak A^{(k)}(A,\xi _1,\ldots ,
\xi _k) &:=\int _N\overline{TQ(\bar i_A^{\ast}\gamma ,\bar i_A
^{\ast}\gamma _0)}^{(k,2m-k-1)}\vert _{u=id_{\Cal G}} (\xi _1,\ldots ,
\xi _k) \\ &=(i_{Z_{\xi _1}^{\Cal A}}\cdots i_{Z_{\xi _k}^{\Cal A}}\int _N
\overline{TQ(\gamma ,\gamma _0)}^{(k,2m-k-1)})(A).\endsplit\tag3.17$$
Regarding $\tilde\frak A^{(k)}$ as a map $\Cal A\rightarrow\wedge ^kLie
\Cal G^{\ast}$, one has $\tilde\frak A^{(k)}=\int _N\bar\chi ^{(k,2m-k-1)}$.
Let
$\wedge ^k ad^{\ast}\Cal A=\Cal A\times _{\Cal G}\wedge ^k Lie\Cal G^{\ast}$
denote the bundle over $\Cal M$ associated to $\Cal A\rightarrow\Cal M$
with respect to the coadjoint action then the above proposition implies that
$\tilde\frak A^{(k)}$ induces a section of $\wedge ^kad^{\ast}\Cal A$. This
result has been obtained in a different way in [K1]. Since the Lie
derivative of $\tilde\frak A^{(k)}$ yields
$$(L_{Z_{\xi}^{\Cal A}}\tilde\frak A^{(k)})(A)(\xi _1,\ldots ,\xi _k)=
\sum _{i=1}^{k}\tilde\frak A^{(k)}(A,\xi _1,\ldots ,[\xi ,\xi _i],\ldots ,
\xi _k),\tag3.18$$
one finally obtains, using (3.7)
$$\multline
i_{Z_{\xi _1}^{\Cal A}}\cdots i_{Z_{\xi _k}^{\Cal A}}\int _N\overline{Q(
\Omega _0)}^{(k,2m-k)}\\ =-\sum _{1\leq i<j\leq k}(-1)^{(i+j)}
\tilde\frak A^{(k)}(A,[\xi _i,\xi _j],\xi _1\ldots ,\hat\xi _i,\ldots ,\hat
\xi _j,\ldots ,\xi _k),\endmultline\tag3.19$$
where the caret denotes omission of the corresponding element.\par
In the remainder of this section we want to investigate the relation between
consistent and covariant terms in more detail. In Sec. 2 the consistent
anomalies have been identified with the family of forms $TQ(\bar i_A^{\ast}
\gamma )^{(k,2m-k-1)}$. Because of the identity
$$TQ(\gamma ,\gamma _0)=TQ(\gamma )-TQ(\gamma _0)+d_{\Cal A\times P}S_Q(
\gamma ,\gamma _0)\tag3.20$$
the general counterterm relating the non-integrated consistent
and covariant anomalies is given by
$$\aligned \lambda (\gamma ,\gamma _0)^{(k,2m-k-1)}:= & TQ(\gamma _0)^{(k,
2m-k-1)}-d_{\Cal A}S_Q(\gamma ,\gamma _0)^{(k-1,2m-k-1)}\\ &+(-1)^{k+1}d_P
S_Q(\gamma ,\gamma _0)^{(k,2m-k-2)}.\endaligned\tag3.21$$
Up to a $d_P$ exact form, the counterterm for $k=1$ is given by
$$TQ(\gamma _0)^{(1,2m-2)}=m(m-1)\int _0^1dt\ tQ(A,d_{\Cal A}A,(F_A)_t,
\ldots ,(F_A)_t),\tag3.22$$
where (3.4) has been used. This is the expression for the non-integrated
Bardeen-Zumino functional [BZ] in dimension $2m-2$.
In order to find the corresponding expressions for the integrated anomalies
we shall use the background connection $\gamma ^a$ of (2.22). The secondary
characteristic forms corresponding to the connections $\gamma$, $\gamma _0$
and $\gamma _a$ fulfill the following general identity
("triangle formula") [MSZ]
$$TQ(\gamma ,\gamma _a)+TQ(\gamma _a,\gamma _0)+TQ(\gamma _0,\gamma )=
d_{\Cal A\times P}SQ(\gamma ^a,\gamma ,\gamma _0),\tag3.23$$
with
$$SQ(\gamma ^a,\gamma ,\gamma _0)=m(m-1)\int _{t_1+t_2\leqq 1}dt_1dt_2\ Q(
\gamma
-\gamma _0,\gamma ^a-\gamma _0,\Omega _{t_1t_2},\ldots ,\Omega _{t_1t_2}),
\tag3.24$$
where $\Omega _{t_1t_2}$ is the curvature of $\gamma _0+t_1(\gamma -
\gamma _0)+t_2(\gamma ^a-\gamma _0)$. $SQ(\gamma ^a,\gamma ,\gamma _0)$
is a basic $(2m-2)$ form on $\Cal A\times P$. Then the counterterm with
background connection becomes
$$\aligned\lambda (\gamma ,\gamma _0,\gamma _a)^{(k,2m-k-1)}:= & TQ(
\gamma _0,\gamma _a)
^{(k,2m-k-1)}-d_{\Cal A}SQ(\gamma ^a,\gamma ,\gamma _0)^{(k-1,2m-k-1)}\\ &+
(-1)^{k+1}d_PSQ(\gamma ^a,\gamma ,\gamma _0)^{(k,2m-k-2)},\endaligned
\tag3.25$$
which is a basic form with projection $\overline{\lambda (\gamma ,\gamma _0,
\gamma _a)}^{(k,2m-k-1)}\in\Omega ^{(k,2m-k-1)}(\Cal A\times M)$.\par
Let $N$ be a $(2m-k-1)$-dimensional submanifold of $M$ without boundary then
the integrated consistent and covariant anomaly of
ghost degree $k$ are related by a generalized Bardeen-Zumino relation
$$\tilde\frak A^{(k)}(\xi _1,\ldots ,\xi _k) =\frak A^{(k)}(\xi _1,\ldots ,
\xi _k)-i_{Z_{\xi _1}^{\Cal A}}\cdots i_{Z_{\xi _k}^{\Cal A}}\int _N
\overline{\lambda (\gamma ,\gamma _0,\gamma _a)}^{(k,
2m-k-1)}.\tag3.26$$
Since $Q(\Omega _0 )=Q(\Omega _0)-Q(\Omega ^a)=d_{\Cal A\times P}TQ(
\gamma _0,\gamma ^a)$, one obtains using (3.25)
$$\int _N\overline{Q(\Omega _0)}^{(k+1,2m-k-1)}=d_{\Cal A}\int _N\overline{
\lambda (\gamma ,\gamma _0,\gamma ^a)}^{(k,2m-k-1)}.\tag3.27$$
In order to make contact with the result of [BC1]
we shall choose $n=\text{dim}M=2m-2$. This implies
$$0=Q(\bar i_A^{\ast}\Omega _0)^{(1,n+1)}=d_PTQ(\bar i_A^{\ast}\gamma ,
\bar i_A^{\ast}
\gamma _0)^{(1,n)}=d_P(TQ(\bar i_A^{\ast}\gamma )^{(1,n)}-TQ(\bar i_A^{\ast}
\gamma _0)^{(1,n)})\tag3.28$$
and by (3.21) we obtain
$$[TQ(\bar i_A^{\ast}\gamma ,\bar i_A^{\ast}\gamma _0)^{(1,n)}]=
[TQ(\bar i_A^{\ast}\gamma )^{(1,n)}-TQ(\bar i_A^{\ast}
\gamma _0)^{(1,n)}]\in H_{d_P}^{(1,n)}(\Cal G\times P,\Bbb R).\tag3.29$$
Actually, $[\overline{TQ(\bar i_A^{\ast}\gamma ,\bar i_A^{\ast}\gamma _0)}
^{(1,n)}]
\in H_{d_M}^{(1,n)}(\Cal G\times M,\Bbb R)$ because the transgression form
is basic.\par
Since the form $TQ(\bar i_A^{\ast}\gamma ,\bar i_A^{\ast}\gamma _0)^{(1,n)}$
is local in the sense of [BC1], i.e. polynomial in the fields $A$
and
$F_A$ and linear in the ghost, it finally gives a certain class in the local
de-Rham cohomology $H_{d_P}^{(1,n)}(\Cal G\times P,\Bbb R)_{
\text{loc}}$.
So we have recovered the result of [BC1] as a direct consequence of the
generalized descent equations.
\bigskip
\bigskip
\head 4. BRS, anti-BRS transformations and the covariance condition\endhead
\bigskip
In [Ts] and [Z] a covariance condition for covariant anomalies has been
formulated in terms of the BRS/anti-BRS algebra. The aim of this section is
to analyze the geometrical meaning of this condition. The first step will
be the construction of a geometrical realization of the BRS/anti-BRS
multiplet.\par
Let us consider the principal $G$-bundle $\Cal A\times\Cal G\times P@>\tilde
\pi >>\Cal A\times\Cal G\times M$, denoted by $\Cal P$, with principal
action $\tilde R_g(A,u,p)=(A,u,pg)$ and
projection $\tilde\pi$. The map $\Cal R^{\Cal A}\times id_P$ is a $G$
bundle homomorphism which makes the following diagram commutative
$$\CD\Cal A\times\Cal G\times P @>\Cal R^{\Cal A}\times id_P>>\Cal A\times P
\\ @V\tilde\pi VV @V\pi VV \\ \Cal A\times\Cal G\times M @>\Cal R^{\Cal A}
\times id_M >> \Cal A\times M\endCD\tag4.1$$
In fact, $(\Cal R^{\Cal A}\times id_M)^{\ast}\Cal Q\cong\Cal P$.
Here $\Omega (\Cal A\times\Cal G\times P)=\oplus _{i,j,k\geq 0}\Omega ^{(i,j,
k)}(\Cal A\times\Cal G\times P)$ admits a tripel grading. Hence we write
$d_{\Cal A\times\Cal G\times P}=d_{\Cal A}+\hat d_{\Cal G}+\hat d_P$, where
$$\aligned &\hat d_{\Cal G}:=(-1)^id_{\Cal G}\colon\Omega ^{(i,j,k)}(\Cal A
\times\Cal G\times P)\rightarrow\Omega ^{(i,j+1,k)}(\Cal A\times\Cal G
\times P)\\ &\hat d_P:=(-1)^{i+j}d_P\colon\Omega ^{(i,j,k)}(\Cal A\times
\Cal G\times P)\rightarrow\Omega ^{(i,j,k+1)}(\Cal A\times\Cal G\times P).
\endaligned\tag4.2$$
We consider the pullback connection $\tilde\gamma :=(\Cal R^{\Cal A}
\times id_P)^{\ast}\gamma$, where $\gamma$ is the connection of
(2.7). The components of $\tilde\gamma$ with respect to the product structure
of $\Omega (\Cal A\times\Cal G\times P)$ are given by
$$\aligned &\tilde\gamma _{(A,v,p)}^{(1,0,0)}(\tau _A)=(((\Cal R_v^{\Cal A})
^{\ast}\alpha )_A)(\tau _A)(p)=ad(\hat v(p)^{-1})(\alpha _A(\tau _A))(p)\\ &
\tilde\gamma _{(A,v,p)}^{(0,1,0)}(\Cal Y_v)=(\Theta _v(\Cal Y_v))(p)\\ &
\tilde\gamma _{(A,v,p)}^{(0,0,1)}(X_p)=(v^{\ast}A)_p(X_p),\endaligned
\tag4.3$$
where $\tau _A\in T_A\Cal A$, $\Cal Y_v\in T_v\Cal G$ and $X_p\in T_pP$.
It is not difficult to calculate the curvature $\tilde\Omega$ of $\tilde
\gamma$. In fact,
$$\split\tilde\Omega = & d_{\Cal A\times\Cal G\times P}\tilde\gamma +\frac{
1}{2}[\tilde\gamma ,\tilde\gamma ]\\= &(\Cal R_v^{\Cal A})^{\ast}\Cal F+
d_{\Cal G}\Theta +\frac{1}{2}[\Theta ,\Theta ]+v^{\ast}F_A+
d_{\Cal A}(v^{\ast}A)-d_{v^{\ast}A}(\Cal R_v^{\Cal A})^{\ast}\alpha\\ & +
d_{\Cal G}((\Cal R_v^{\Cal A})^{\ast}\alpha )-[(\Cal R_v^{
\Cal A})^{\ast}
\alpha ,\Theta ]+d_{\Cal G}(v^{\ast}A)-d_{v^{\ast}A}\Theta .\endsplit
\tag4.4$$
Let $\tau _A\in T_A\Cal A$ be a fixed
tangent vector then $v\mapsto (\Cal R_v^{\Cal A\ \ast}\alpha )_A(\tau _A)$
can be viewed as a $Lie\Cal G$-valued function on $\Cal G$. The vector
$\Cal Y_v=\Cal Y_{\Theta _v(\Cal Y_v)}^{\text{left}}(v)\in T_v\Cal
G$ generates the flow $Fl_t^{\Theta _v(\Cal Y_v)}(v)=\hat v\cdot expt
\Theta _v(\Cal Y_v)$. So we find
$$\split (d_{\Cal G}(((\Cal R_.^{\Cal A})^{\ast}\alpha )_A(\tau _A))_v(
\Cal Y_v)(p) &=\frac {d}{dt}\vert _{t=0}\ ad(exp(-t\Theta _v(\Cal Y_v)(p)))
\ ad(\hat v^{-1}(p))(\alpha _A(\tau _A))(p)\\ &=[((\Cal R_v^{\Cal A})^{\ast}
\alpha )_A(\tau _A),\Theta _v(\Cal Y_v)](p).\endsplit\tag4.5$$
Inserting (2.14) and (4.5) into (4.4) leads to
\proclaim{4.6 Proposition} The components of the curvature $\tilde\Omega$
of the connection $\tilde\gamma$ are given by
$$\aligned & \tilde\Omega _{(A,v,p)}^{(2,0,0)}(\tau _A^1,\tau _A^2)=((
(\Cal R_v^{\Cal A})^{\ast}\Cal F)_A(\tau _A^1,\tau _A^2))(p)=ad(\hat v(p)
^{-1})(\Cal F_A(\tau _A^1,\tau _A^2))(p)\\ &
\tilde\Omega _{(A,v,p)}^{(0,0,2)}(X_p^1,X_p^2)=(v^{\ast}F_A)_p(X_p^1,
X_p^2)\\ & \tilde\Omega _{(A,v,p)}^{(1,0,1)}(\tau _A,X_p)=(v^{\ast}
\tau _A^h)_p(X_p)\\ &
\tilde\Omega ^{(0,2,0)}=\tilde\Omega ^{(1,1,0)}=
\tilde\Omega ^{(0,1,1)}=0.\endaligned$$ \endproclaim
Let us define a free $\Cal G$ action on $\Cal A\times\Cal G\times P$ by
$$\tilde\Cal R_u(A,v,p):=(\Cal R_u^{\Cal A}\times Ad(u^{-1})\times u^{-1})
(A,v,p)=(u^{\ast}A,u^{-1}vu,u^{-1}(p)),\tag4.7$$
then it is easily verified that $(\Cal R^{\Cal A}\times id_P)\circ\tilde
\Cal R_u
=\Cal R_u\circ (\Cal R^{\Cal A}\times id_P)$ holds. Since $\gamma$ is
$\Cal G$ invariant we find $\tilde
\Cal R_u\tilde\gamma =\tilde\gamma$. The quotient $(\Cal A\times\Cal G\times
P)/\Cal G$ admits a
canonical free $G$-action. The induced $\Cal G$-action on $\Cal A\times
\Cal G\times M$ is given by $(A,v,x)\mapsto (u^{\ast}A,u^{-1}vu,x)$.
Let $Ad\Cal A=\Cal A\times _{\Cal G}\Cal G$ denote the adjoint bundle,
associated to the principal bundle $\Cal A$ via the adjoint action of $\Cal
G$ onto itself. Then one can prove $(\Cal A\times\Cal G\times M)/\Cal G
\cong Ad\Cal A\times M$ and thus $(\Cal A\times\Cal G\times P)/\Cal G
\rightarrow Ad\Cal A\times M$ becomes a principal $G$-bundle, denoted by
$\hat\Cal P$. In summary
there exists the following commutative diagram of principal bundles
$$\CD\Cal A\times\Cal G\times P @>r>>(\Cal A\times\Cal G\times P)/\Cal G\\
@V\tilde\pi VV @V\tilde\pi ^{\prime}VV \\ \Cal A\times\Cal G\times M
@>\hat r>> Ad\Cal A\times M\endCD\tag4.8$$
The vertical subspaces $V_{(A,v,p)}^r(\Cal A\times\Cal G\times P)$ of the
principal bundle $\Cal A\times\Cal G\times P\rightarrow (\Cal A\times\Cal G
\times P)/\Cal G$ are given by
$$V_{(A,v,p)}^r(\Cal A\times\Cal G\times P)=\lbrace (d_A\xi ,\Cal Y_{\xi}^{
\text{left}}(v)-\Cal Y_{\xi}^{\text{right}}(v),-Z_{\xi}(p))\vert\xi\in
\Omega ^0(M,adP)\rbrace ,\tag4.9$$
where $\Cal Y_{\xi}^{\text{right}}$ denotes the right invariant vector
field on $\Cal G$ induced by $\xi$. It is easy to show that $\tilde
\gamma\vert _{V^r(\Cal A\times\Cal G\times P)}=0$ and so we end up with
\proclaim{4.10 Proposition} The connection $\tilde\gamma =(\Cal R^{\Cal A}
\times id_P)^{\ast}\gamma $ descends to a well defined connection on the
$G$-bundle $(\Cal A\times\Cal G\times P)/\Cal G\rightarrow Ad\Cal A\times M$.
\endproclaim
It is clear that $\hat r^{\ast}\hat\Cal P\cong\Cal P$. So $\Cal P$ with
connection $\tilde\gamma$ can serve as corresponding generalization of
$\Cal Q$ with universal connection $\gamma$.\par
Let $\tilde i_A:\Cal G\times\Cal G\times P\hookrightarrow \Cal A\times\Cal G
\times P, (u,v,p)\mapsto (u^{\ast}A,v,p)$ be the embedding of the gauge
orbit through $A\in\Cal A$.
The components of the restriction of $\tilde\gamma$ are given by
$$\aligned (\tilde i_A^{\ast}\tilde\gamma )_{(u,v,p)}^{(1,0,0)}(\Cal Y_u) &=
ad(\hat v(p)^{-1})\Theta _u(\Cal Y_u)(p)\\
(\tilde i_A^{\ast}\tilde\gamma )_{(u,v,p)}^{(0,1,0)}(\Cal Y_v) &=\Theta _v
(\Cal Y_v)(p)\\(\tilde i_A^{\ast}\tilde\gamma )_{(u,v,p)}^{(0,0,1)}(X_p) &=
((uv)^{\ast}A)_p(X_p).\endaligned\tag4.11$$
Since $\tilde\Omega$ is horizontal with respect
to the principal fibration $\Cal A\rightarrow\Cal M$, a generalization of
the Russian formula holds, namely
$$(\tilde i_A^{\ast}\tilde\Omega )_{(u,v,p)}((\Cal Y_u^1,\Cal Y_v^{\prime 1},
X_p^1),(\Cal Y_u^2,\Cal Y_v^{\prime 2},X_p^2))=((uv)^{\ast}F_A)_p(X_p^1,
X_p^2).\tag4.12$$
Now we define $\tilde A:=\tilde i_A^{\ast}
\tilde\gamma ^{(0,0,1)}$, $\bar\Theta :=\tilde i_A^{\ast}\tilde\gamma
^{(1,0,0)}$ and $\tilde F_A:=\tilde i_A^{\ast}\tilde\Omega$. Because of
(4.12) the components of $\tilde i_A^{\ast}\tilde\gamma$ fulfill
the relations
$$\alignat2 & d_{\Cal G}^{(1)}\tilde A -d_{\tilde A}\bar\Theta =0,
&& \qquad d_{\Cal G}^{(2)}\tilde A -d_{\tilde A}\Theta =0\\
& d_{\Cal G}^{(1)}\bar\Theta +\frac{1}{2}[\bar\Theta ,\bar\Theta ]=0, &&
\qquad d_{\Cal G}^{(2)}\Theta +\frac{1}{2}[\Theta ,\Theta ]=0\tag4.13\\
& d_{\Cal G}^{(2)}\bar\Theta +[\bar\Theta ,
\Theta ]=0, && \qquad d_{\Cal G}^{(1)}\Theta =0,\endalignat$$
where $d_{\Cal G}^{(1)}$ and $d_{\Cal G}^{(2)}$ denote the exterior
derivatives with
respect to the first and second factor in the product $\Cal G\times\Cal G
\times P$. Furthermore the Bianchi identity gives
$$d_{\Cal G}^{(1)}\tilde F_A+[\bar\Theta ,\tilde F_A]=0\qquad d_{\Cal G}
^{(2)}\tilde F_A+[\Theta ,\tilde F_A]=0.\tag4.14$$
In comparison with [Ts, Z] we can identify $d_{\Cal G}^{(1)}$ with the
anti-BRS operator and $d_{\Cal G}^{(2)}$ with the BRS operator. Moreover
$\Theta$ is identified with the ghost and $\bar\Theta$ plays the role of
the anti-ghost field. Thus we have shown that the BRS/anti-BRS multiplet
can be geometrically realized on the $G$-bundle $\Cal G\times\Cal G
\times P\rightarrow\Cal G\times\Cal G\times M$.\par
Let $Q\in I^m(G)$
and define the $2m$ form $Q(\tilde\Omega )$. The transgression formula
$Q(\tilde\Omega )=
d_{\Cal A\times\Cal G\times P}TQ(\tilde\gamma )$ gives rise to general
descent equations. However, by restricting to the gauge orbit through $A\in
\Cal A$ this system can be separated into a set, which corresponds
to the cohomology of $d_{\Cal G}^{(1)}$ and $d_{\Cal G}^{(2)}$ respectively
$$\split Q(\tilde i_A^{\ast}\tilde\Omega )^{(0,0,2m)} &=d_PTQ(\tilde i_A^{
\ast}\tilde\gamma )^{(0,0,2m-1)}\\ 0 &=d_{\Cal G}^{(2)}TQ(\tilde i_A^{
\ast}\tilde\gamma )^{(0,k-1,2m-k)}+\hat d_PTQ(\tilde i_A^{
\ast}\tilde\gamma )^{(0,k,2m-1-k)}\qquad 1\leqq k\leqq 2m\\
0 &=d_{\Cal G}^{(1)}TQ(\tilde i_A^{
\ast}\tilde\gamma )^{(k-1,0,2m-k)}+\hat d_PTQ(\tilde i_A^{
\ast}\tilde\gamma )^{(k,0,2m-1-k)}\qquad 1\leqq k\leqq 2m.\endsplit\tag4.15$$
There is also a mixed set of equations for $1\leqq k\leqq 2m$
$$\split 0 &=d_{\Cal G}^{(1)}TQ(\tilde i_A^{\ast}\tilde\gamma )^{(0,k-1,
2m-k)}+\hat d_{\Cal G}^{(2)}TQ(\tilde i_A^{\ast}\tilde\gamma )^{(1,k-2,
2m-k)}+\hat d_PTQ(\tilde i_A^{\ast}\tilde\gamma )^{(1,k-1,2m-k-1)}
\\ 0 &=d_{\Cal G}^{(1)}TQ(\tilde i_A^{\ast}\tilde\gamma )^{(k-2,1,
2m-k)}+\hat d_{\Cal G}^{(2)}TQ(\tilde i_A^{\ast}\tilde\gamma )^{(k-1,0,2m-k)}
+\hat d_PTQ(\tilde i_A^{\ast}\tilde\gamma )^{(k-1,1,2m-k-1)}.
\endsplit\tag4.16$$
Since the components of $\tilde\Omega _t=t\tilde\Omega +
\frac{t^2-t}{2}[\tilde\gamma ,\tilde\gamma ]$ read
$$\aligned &\tilde\Omega _{t\ (A,v,p)}^{\ \ (2,0,0)}(\tau _A^1,\tau _A^2)=
((\Cal R_v^{\Cal A})^{\ast}\Cal F_t)_A(\tau _A^1,\tau _A^2)(p) \\ &
\tilde\Omega _{t\ (A,v,p)}^{\ \ (0,2,0)}(\Cal Y_v^1,\Cal Y_v^2) =
(\frac{t^2-t}{2}[\Theta _v,\Theta _v])(\Cal Y_v^1,\Cal Y_v^2)(p)
\\ &\tilde\Omega _{t\ (A,v,p)}^{\ \ (0,0,2)}(X_p^1,X_p^2)=(v^{\ast}(F_A)_t)_p
(X_p^1,X_p^2) \\
&\tilde\Omega _{t\ (A,v,p)}^{\ \ (1,1,0)}(\tau _A,\Cal Y_v) =(t^2-t)[(
(\Cal R_v^{\Cal A})^{\ast}\alpha )_A,\Theta _v](\tau _A,\Cal Y_v)(p)\\
&\tilde\Omega _{t\ (A,v,p)}^{\
\ (0,1,1)}(\Cal Y_v,X_p)=(t^2-t)[\Theta _v,v^{\ast}A](\Cal Y_v,X_p)\\
&\tilde\Omega _{t\ (A,v,p)}^{\ \ (1,0,1)}(\tau _A,X_p)= (t\tilde
\Omega _{(A,v,p)}
^{(1,0,1)}+(t^2-t)[((\Cal R_v^{\Cal A})^{\ast}\alpha )_A,u^{\ast}A])
(\tau _A,X_p),\endaligned\tag4.17$$
we obtain, in view of (2.17)
$$TQ(\tilde i_A^{\ast}\tilde\gamma )_{(u,id_{\Cal G},p)}^{(k,0,2m-k-1)}=
TQ(\bar i_A^{\ast}\gamma )_{(u,p)}^{(k,2m-k-1)}.\tag4.18$$
So we have recovered the usual expression for the consistent anomalies.\par
In order to formulate covariant descent equations in our present setup, we
define the following connection
$$\eta _{(A,v,p)}(\tau _A,\Cal Y_v,X_p)=(v^{\ast}A)_p(X_p)+(\Theta _v(
\Cal Y_v))(p)\tag4.19$$
on the bundle $\Cal A\times\Cal G\times P\rightarrow\Cal A
\times\Cal G\times M$. However, $\eta$ is not the pullback of $\gamma _0$.
It is $\Cal G$ invariant since
$$\split (\tilde\Cal R_u^{\ast}\eta )_{(A,v,p)} &=\eta _{(u^{\ast}A,
Ad(u^{-1})
v, u^{-1}(p))}(u^{\ast}\tau _A,T_vAd(u^{-1})\Cal Y_v,T_pu^{-1}X_p)\\
&=((vu)^{\ast}A)_{u^{-1}(p)}(T_pu^{-1}X_p)+ad(\hat u(u^{-1}(p)))(
\Theta _v(\Cal Y_v))(u^{-1}(p))\\
&=\eta _{(A,v,p)}(\tau _A,\Cal Y_v,X_p).\endsplit\tag4.20$$
The curvature $\Omega _{\eta}$ of $\eta$ is given by
$$\split\Omega _{\eta} &=d_{\Cal A\times\Cal G\times P}\ \eta +
\frac{1}{2}[
\eta ,\eta ]\\ &=v^{\ast}F_A+\hat d_{\Cal G}v^{\ast}A-d_{v^{\ast}
A}\Theta _v+d_{\Cal A}v^{\ast}A\\ &= v^{\ast}F_A+d_{\Cal A}v^{\ast}A,
\endsplit\tag4.21$$
where the BRS relations (2.14) have been used. In summary, the components of
$\Omega _{\eta}$ read
$$\aligned &\Omega _{\eta}^{\ (2,0,0)}=\Omega _{\eta}^{\ (0,2,0)}=
\Omega _{\eta}^{\ (1,1,0)}=\Omega _{\eta}^{\ (0,1,1)}=0\\
&\Omega _{\eta\ (A,v,p)}^{\ \ (0,0,2)}(X_p^1,X_p^2)=(v^{\ast}F_A)_p(X_p^1,
X_p^2)\\ &
\Omega _{\eta\ (A,v,p)}^{\ \ (1,0,1)}(\tau _A,X_p)=(v^{\ast}\tau _A)_p(X_p).
\endaligned\tag4.22$$
The corresponding transgression formula is $Q(\Omega _{\eta})=d_{\Cal A\times
\Cal G\times P}TQ(\eta )$. Using (4.13), the components of $(\Omega _{
\eta})_t=t\Omega _{\eta}+\frac{t^2-t}{2}[\eta ,\eta ]$ read
$$\aligned &(\Omega _{\eta})_t^{\ \ (2,0,0)}=(\Omega _{\eta})_t^{\ \ (1,1,0)}
=0\\
&(\Omega _{\eta})_{t\ (A,v,p)}^{\ \ (0,0,2)}(X_p^1,X_p^2)=(v^{\ast}(F_A)_t)_p
(X_p^1,X_p^2)\\ &(\Omega _{
\eta})_{t\ (A,v,p)}^{\ \ (1,0,1)}(\tau _A,X_p)=td_{\Cal A}(v^{
\ast}A)(\tau _A,X_p)=t(v^{\ast}\tau _A)_p(X_p)\\
&(\Omega _{\eta})_{t\ (A,v,p)}^{\ \ (0,2,0)}(\Cal Y_v
^1,\Cal Y_v^2)=\frac{t^2-t}{2}[
\Theta _v,\Theta _v](\Cal Y_v^1,\Cal Y_v^2)(p)
\\ &(\Omega _{\eta})_{t\ (A,v,p)}^{\ \ (0,1,1)}(\Cal Y_v,X_p)=
(t^2-t)[\Theta _v,v^{\ast}A](\Cal Y_v,X_p).\endaligned\tag4.23$$
Hence we find the important relation
$$TQ(\tilde i_A^{\ast}\eta )_{(u,id_{\Cal G},p)}^{(j,0,2m-j-1)}=TQ(
\bar i_A^{\ast}\gamma _0)_{(u,p)}^{(j,2m-j-1)}.\tag4.24$$
Moreover (4.6) and (4.22) imply
$$\aligned Q(\tilde\Omega )^{(j,k-j,2m-k)} &=0,\qquad j\neq k\\
Q(\Omega _{\eta})^{(j,k-j,2m-k)} &=0,\qquad j\neq k.\endaligned\tag4.25$$
The transgression formula $Q(\tilde\Omega )-Q(\Omega _{\eta})=d_{\Cal A\times
\Cal G\times P}TQ(\tilde\gamma ,\eta )$ leads to the descent equations
$$\split (Q(\tilde\Omega )-Q(\Omega _{\eta}))^{(j,k-j,2m-k)}=&
d_{\Cal A}TQ(\tilde\gamma ,\eta )^{(j-1,k-j,2m-k)}+\hat d_{
\Cal G}TQ(\tilde\gamma ,\eta )^{(j,k-j-1,2m-k)}\\ & +\hat d_PTQ(
\tilde\gamma ,\eta )^{(j,k-j,2m-k-1)},\endsplit\tag4.26$$
where $0\leq j,k\leq 2m$.
Furthermore (4.20) implies $(\Cal R_u^{\Cal A}\times id_{\Cal G}\times id_P)
^{\ast}TQ(\tilde\gamma ,\eta )=TQ(\tilde\gamma ,\eta)$. The components of
the curvature $\tilde\Omega _t^{\prime}$ corresponding to the connection
$t\tilde\gamma +(1-t)\eta$ are
$$\aligned &\tilde\Omega _t^{\prime\ (0,2,0)}=\tilde\Omega _t^{\prime\ (0,1,
1)}=\tilde\Omega
_t^{\prime\ (1,1,0)}=0\\ &\tilde\Omega _{t\ (A,v,p)}^{\prime\ (2,0,0)}
(\tau _A^1,\tau _A^2)
=((\Cal R_v^{\Cal A})^{\ast}\Cal F_t)_A(\tau _A^1,\tau _A^2)(p)\\
&\tilde\Omega _{t\ (A,v,p)}^{\prime\ (0,0,2)}(X_p^1,X_p^2)=(v^{\ast}F_A)_p
(X_p^1,X_p^2)\\ &\tilde\Omega _{t\ (A,v,p)}^{\prime\ (1,0,1)}(\tau _A,X_p)
=(d_{\Cal A}(v^{\ast}A)-td_{v^{
\ast}A}((\Cal R_v^{\Cal A})^{\ast}\alpha _A))(\tau _A,X_p),\endaligned
\tag4.27$$
where we have used the BRS/anti-BRS relations (4.13). It is obvious from
(3.8) and (4.27) that
$$TQ(\tilde i_A^{\ast}\tilde\gamma ,\tilde i_A^{\ast}\eta )_{(u,id_{
\Cal G},p)}^{(j,0,2m-j-1)}=TQ(\bar i_A^{\ast}\gamma ,\bar i_A^{\ast}\gamma _0
)_{(u,p)}^{(j,2m-j-1)}\tag4.28$$
holds. Generally, in consequence of (4.27) one finds
$$TQ(\tilde\gamma ,\eta )^{(j,k-j,2m-k-1)}=0\qquad j\neq k.\tag4.29$$
Choose $k=j+1$, then the descent equation (4.26) reduces to
$$d_{\Cal G}TQ(\tilde\gamma ,\eta )^{(j,0,2m-j-1)}=0\tag4.30$$
and so one ends up with
$$d_{\Cal G}^{(2)}TQ(\tilde i_A^{\ast}\tilde\gamma ,\tilde i_A^{\ast}\eta
)^{(j,0,2m-j-1)}=0,\tag4.31$$
which in [AAG] is called the strong covariance condition for the
non-integrated covariant
anomalies. However, this result could have been also
obtained by
a direct computation using the BRS/anti-BRS transformations (4.13), (4.14)
and the property of $Q$ being ad-invariant. Thus we have shown that the
solutions of the covariant descent equations automatically satisfy the
covariance condition. \par
In view of
$$TQ(\tilde\gamma ,\eta )=TQ(\tilde\gamma )-TQ(\eta )+d_{\Cal A
\times\Cal G\times P}S_Q(\tilde\gamma ,\eta ),\tag4.32$$
and following the procedure of Sec. 2,
the counterterm relating consistent and covariant anomalies may be defined by
$$\split\lambda (\tilde\gamma ,\eta )^{(j,0,2m-j-1)}:= & TQ(\eta )^{(j,0,
2m-j-1)}-d_{\Cal A}S_Q(\tilde\gamma ,\eta )^{(j-1,0,2m-j-1)}\\
&+(-1)^{j+1}d_PS_Q(\tilde\gamma ,\eta )^{(j,0,2m-j-2)}.\endsplit\tag4.33$$
Note that $S_Q(\tilde\gamma ,\eta )^{(0,k,2m-k-2)}=0$, since $(\tilde
\gamma -\eta )\in\Omega ^{(1,0,0)}(\Cal A\times\Cal G\times P,
\frak g)$. For sake of completeness we want to write the covariance
condition (4.31) in terms of the consistent anomaly and the corresponding
counterterm. Let us assume, for simplicity, that $P=M\times G$ is
trivial and let $\sigma$ denote a global section of $\Cal G\times\Cal G
\times P\rightarrow\Cal G\times\Cal G\times M$.
Using (4.31) and (4.33), we find
$$\split 0 = d_{\Cal G}^{(2)}\int _N\sigma ^{\ast}[ & TQ(\tilde i_A^{\ast}
\tilde\gamma )^{(j,0,
2m-j-1)}-(TQ(\tilde i_A^{\ast}\eta )^{(j,0,2m-j-1)}\\ &-d_{\Cal G}
^{(1)}S_Q(\tilde i_A^{\ast}\tilde\gamma ,\tilde i_A^{\ast}\eta )^{(j-1,0,
2m-j-1})],\endsplit\tag4.34$$
where $N$ is a $(2m-j-1)$ dimensional submanifold of $M$. For
$j=1$, (4.34) reduces to the well known covariance condition for
the integrated covariant current anomaly
$$d_{\Cal G}^{(2)}\int _M\sigma ^{\ast}[TQ(\tilde i_A^{\ast}\tilde
\gamma )^{(1,0,
2m-2)}-TQ(\tilde i_A^{\ast}\eta )^{(1,0,2m-2)}]=0.\tag4.35$$
It has been argued in Ref. 6 that two homotopy operators have to be
introduced to generate the covariant anomalies.
In our approach this assertion is nothing but the fact that the
counterterm $\lambda (\tilde i_A^{\ast}\tilde\gamma ,
\tilde i_A^{\ast}\eta )^{(j,0,2m-j-1)}$ consists of two different
types of local polynomials, namely $TQ(\tilde i_A^{
\ast}\eta )^{(j,0,2m-j-1)}$ and $S_Q(\tilde i_A^{\ast}\tilde\gamma ,\tilde
i_A^{\ast}\eta )^{(j,0,2m-j-1)}$. However, if $P\rightarrow M$ is non-trivial
the generalization of the background formalism discussed previously is
straightforward.\par
Using (4.32), one finds from (4.26)
$$\split Q(\tilde i_A^{\ast}\Omega _{\eta})^{(1,0,2m-1)} & =d_PTQ(\tilde
i_A^{\ast}
\tilde\gamma ,\tilde i_A^{\ast}\eta )^{(1,0,2m-2)}\\ &=d_P(TQ(\tilde
i_A^{\ast}\tilde\gamma )^{(1,0,2m-2)}-TQ(\tilde i_A^{\ast}\eta )^{(1,0,
2m-2)}).\endsplit\tag4.36$$
If $n=dimM=2m-2$, the left hand side of (4.36) vanishes
and so the non-integrated covariant current anomaly induces the local de-Rham
cohomology class
$$[TQ(\tilde i_A^{\ast}\tilde\gamma ,\tilde i_A^{\ast}\eta )^{(1,0,n)}]=
[TQ(\tilde i_A^{\ast}\tilde\gamma )^{(1,0,n)}-TQ(\tilde i_A^{\ast}\eta )
^{(1,0,n)}]\in H_{d_P}^{(1,0,2m-2)}(\Cal G\times\Cal G\times P,\Bbb R)
\tag4.37$$
in complete analogy with the result obtained in the previous section.\par
In summary we have found two conditions
$$\split & d_{\Cal G}^{(2)}TQ(\tilde i_A^{\ast}\tilde\gamma ,\tilde i_A^{
\ast}\eta )^{(1,0,n)} =0\\ & d_PTQ(\tilde i_A^{\ast}\tilde\gamma ,\tilde i_A^{
\ast}\eta )^{(1,0,n)}=0\endsplit\tag4.38$$
for the non-integrated covariant current anomaly. These conditions are a
direct consequence of the covariant descent equations which we have derived
in this paper. Thus
we have shown how the covariance condition for the covariant anomaly is
related with its characterization in terms of local de-Rham cohomology.
\bigskip\bigskip
\head Acknowledgement\endhead
I would like to thank Prof. J. Wess for his kind hospitality at the
LMU of Munich.

\enddocument